\documentclass[a4paper,11pt]{article}
\usepackage[total={6.1in, 8.5in}]{geometry}
\usepackage[utf8]{inputenc}
\usepackage{amssymb}
\usepackage{amsmath}
\usepackage{braket}
\usepackage{hyperref}
\usepackage{authblk}
\usepackage{graphicx}
\usepackage{textcomp}
\usepackage{subcaption}
\usepackage{slashed}
\usepackage{appendix}

\newcommand{\be}{\begin{equation}}
\newcommand{\ee}{\end{equation}}
\newcommand{\ba}{\begin{eqnarray}}
\newcommand{\ea}{\end{eqnarray}}

\title{ Leading Loop Effects in Pseudoscalar-Higgs Portal Dark Matter}

\author{Karim Ghorbani%
  \thanks{\texttt{karim1.ghorbani@gmail.com}}}
\affil{\it Physics Department, Faculty of Sciences, Arak University, \\
\it Arak 38156-8-8349, Iran}

\author{Parsa Hossein Ghorbani%
  \thanks{\texttt{parsa@appliedphysics.org}}}
\affil{\it Applied Physics Inc., Center for Cosmological Research, \\
\it 300 Park Avenue, New York, NY 10022, USA}

\date{}

\begin{document}

\maketitle

\begin{abstract}
We examine a model with a fermionic dark matter candidate having pseudoscalar 
interaction with the standard model particles where its direct 
detection elastic scattering cross section at tree level is highly suppressed. 
We then calculate analytically the leading loop contribution to the spin independent 
scattering cross section. It turns out that these loop effects are sizable 
over a large region of the parameter space. Taking constraints from 
direct detection experiments, the invisible Higgs decay measurements, observed 
DM relic density, we find viable regions which are within reach 
in the future direct detection experiments such as XENONnT. 
\end{abstract}
\maketitle
\section{Introduction}
\label{int}
A nagging question in contemporary modern physics is about the nature of dark matter (DM) 
and its feasible non-gravitational interaction with the standard model (SM) particles. 
This problem is in fact deemed straddling both particle physics and cosmology.

On the cosmology side, precise measurements of the 
Cosmic Microwave Background (CMB) anisotropy not only demonstrate 
the existence of dark matter but also provide us 
with the current dark matter abundance in the universe \cite{Ade:2013zuv,Hinshaw:2012aka}.
On the particle physics side, the dedicated search is to find direct detection (DD) 
of the DM interaction with the ordinary matter via Spin Independent (SI) or Spin Dependent (SD) 
scattering of DM-nucleon in underground experiments like LUX \cite{Akerib:2016vxi}, 
XENON1T \cite{Aprile:2017iyp} and PandaX-II \cite{Cui:2017nnn}. 
Although in these experiments the enticing signal 
is not shown up so far, the upper limit on the DM-matter interaction strength
is provided for a wide range of the DM mass.
Among various candidates for particle DM, the most sought one is the Weakly
Interacting Massive Particle (WIMP). 

Within WIMP paradigm there exist a class of models where SI scattering cross section
is suppressed significantly at leading order in perturbation 
theory, hence the model eludes the experimental upper limits 
in a large region of the parameter space. The interaction type of the WIMP-nucleon
in these models are pseudoscalar or axial vector at tree level 
resulting in momentum or velocity suppressed 
cross section \cite{Fitzpatrick:2012ix}. 
The focus here is on models with pseudoscalar interaction between the DM particles 
and the SM quarks. In this case there are both SI and SD elastic scattering of the DM 
off the nucleon at tree level. Both type of the interactions 
are momentum dependent while the SD cross section gets suppressed much 
stronger than the SI cross section due to an extra momentum transfer factor, $q^2$. 
Thus, in these models taking into account beyond tree level contributions which 
could be leading loop effects or full one-loop effects are essential. 

We recall several earlier works done in this direction with emphasis 
on DM models with a pseudoscalar interaction. 
Leading loop effect on DD cross section is studied in an extended 
two Higgs doublet model in \cite{Arcadi:2017wqi,Sanderson:2018lmj,Abe:2018emu}.
Within various DM simplified models in \cite{Li:2018qip,Herrero-Garcia:2018koq,Hisano:2018bpz} and in a 
singlet-doublet dark matter model in \cite{Han:2018gej} 
the loop induced DD cross sections are investigated. 
Full one-loop contribution to the DM-nucleon scattering cross section
in a Higgs-portal complex scalar DM model can be found in \cite{Azevedo:2018exj}.
In \cite{Ishiwata:2018sdi} direct detection of a pseudo scalar dark matter 
is studied by taking into account higher order corrections both in QCD
and non-QCD parts.

In this work we consider a model with fermionic DM candidate, $\psi$, which interacts
with a pseudoscalar mediator $P$ as $P \bar \psi \gamma^5 \psi$. The pseudoscalar
mediator will be connected to the SM particles via mixing with the SM Higgs with 
an interaction term as $P H^{\dagger}H$. 
In this model the DM-nucleon interaction at tree level is of pseudoscalar type and thus 
its scattering cross section is highly suppressed over 
the entire parameter space. 
The leading loop contribution to the DD scattering cross section being spin independent
is computed and viable regions are found against the direct detection bounds. 
Beside constraints from observed relic density, the invisible Higgs decay limit 
is imposed when it is relevant.

The outline of this article is as follows. 
In Sec.~\ref{model} we recapitulate the pseudoscalar DM model. 
We then present our main results concerning the direct detection of the DM
including analytical formula for the DD cross section and numerical analysis 
in Sec.~\ref{DD}. Finally we finish with a conclusion.

\section{The Pseudoscalar Model} 
\label{model}
The model we consider in this research as a renormalizable extension to the SM, 
consists of a new 
gauge singlet Dirac fermion as the DM candidate and a new singlet scalar acting  
as a mediator, which connects the fermionic DM to SM particles via the Higgs portal. 
The new physics Lagrangian comes in two parts, 
\begin{equation}
{\cal L}  = {\cal L}_{\text{DM}} +{\cal L}_{\text{scalar}}  \,.
\end{equation}
The first part, ${\cal L}_{\text{DM}}$, introduces a pseudoscalar interaction term as
\begin{equation}
\label{DM-lag}
{\cal L}_{\text{DM}}   = \bar \psi (i {\not}\partial-m_{\text{\text{dm}}})\psi -i g_{d}~ P \bar{\psi} \gamma^5 \psi \,,
\end{equation}
and the second part, ${\cal L}_{\text{scalar}}$, incorporates the 
singlet pseudoscalar and the SM Higgs doublet as
\begin{equation}
\begin{aligned} 
 {\cal L}_{\text{scalar}} {}& =  \frac{1}{2} (\partial_{\mu} P)^2 - \frac{m^{2}}{2} P^2 
                       - \frac{g_3}{6} P^3 
                       -\frac{g_4}{24} P^4 
                       + \mu^{2}_{H} H^{\dagger}H 
                       \\ &
                       - \lambda (H^{\dagger}H)^2 
                       + g_0 P
                       - g_1 P H^{\dagger}H
                       - g_2 P^2 H^{\dagger}H 
                       \,.
\end{aligned}
\end{equation}
The pseudoscalar field is assumed to acquire a zero vacuum expectation value ({\it vev}), 
$\braket{P}  = 0$, while it is known that the SM Higgs 
develops a non-zero {\it vev} where $\braket{H} = v_h = 246$ GeV.
Having chosen $\braket{P}  = 0$, the tadpole coupling $g_0$ is fixed appropriately.
After expanding the Higgs doublet in unitary gauge as $H = ( 0 ~~ v_h+h')^T$, we write 
down the scalar fields in the basis of mass eigenstates $h$ and $s$, in the following expressions
\begin{equation}
 h' = h \cos\theta - s \sin\theta \,, ~~~~~ P = h \sin\theta + s \cos\theta \,. 
\end{equation}
The mixing angle, $\theta$, is induced by the interaction term $P H^{\dagger}H$ and 
is obtained by the relation $\sin2\theta = 2g_1 v_h/(m_h^2-m_s^2)$, in which $m_h = 125$ GeV
and $m_s$ are the physical masses for the Higgs and the singlet scalar, respectively.
The quartic Higgs coupling is modified now and is given in terms of the mixing angle 
and the physical masses of the scalars as $\lambda = (m_h^2 \cos^2\theta + m_s^2 \sin^2\theta)/(2v_h^2)$.
We can pick out as independent free parameters a set of parameters as $\theta$, $g_d$, $g_2$, $g_3$, $g_4$ and $m_s$. 
The coupling $g_1$ is then fixed by the relation $g_1 = \sin2\theta (m_h^2-m_s^2)/(2v_h)$.
Recent study on the DM and the LHC phenomenology 
of this model can be found in \cite{Ghorbani:2014qpa,Baek:2017vzd}
and its electroweak baryogenesis is examined in \cite{Ghorbani:2017jls}.

For DM masses in the range $m_{\text{dm}} < m_h/2$, one can impose constraint on the parameters
$g_d$, $\theta$ and $m_{\text{dm}}$ from invisible Higgs decay measurements 
with Br($h\to$ invisible) $\lesssim 0.24$ \cite{Khachatryan:2016whc}.
Given the invisible Higgs decay process, $h \to \bar \psi \psi$, we find for small 
mixing angle the condition 
$g_d \sin \theta \lesssim 0.16~\text{GeV}^{1/2}/(m_h^2-4m_{\text{dm}}^2)^{1/4}$ \cite{Kgh-MonoHiggs2017}. 

We compute DM relic density numerically over the model parameter space
by applying the program {\tt micrOMEGAs} \cite{Barducci:2016pcb}.
The observed value for the DM relic density used in our numerical 
computations is $\Omega h^2 = 0.1198\pm 0.0015$ \cite{Ade:2015xua}.
The DM production in this model 
is via the popular freeze-out mechanism \cite{Lee:1977ua} in which it is assumed that DM particles 
have been in thermal equilibrium with the SM particles in the early universe. 

\begin{figure}
\hspace{1.9cm}
\includegraphics[width=.5\textwidth,angle =-90]{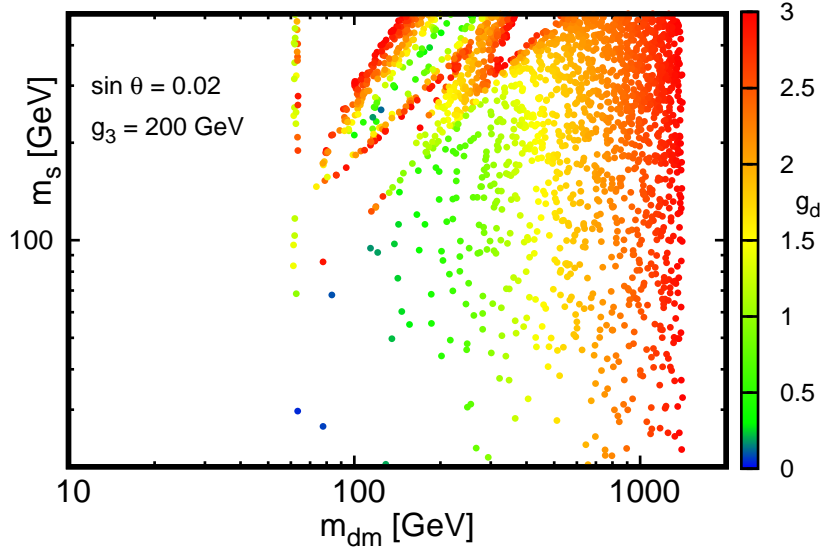}
\caption{The viable region shown in the $m_s-m_{\text{dm}}$ plane respects the restrictions from the observed relic density and the measurements of the invisible Higgs decay.}
\label{relic-space}
\end{figure}

We find the viable region in the parameter space respecting the constraints from observed relic density 
and invisible Higgs decay in Fig.~\ref{relic-space}. The parameters chosen in this computation are
$\sin \theta = 0.02$, $g_{3} = 200$ GeV and $g_2 = 0.1$. It is evident in the plot that regions 
with $m_{\text{dm}} < m_h/2$ are excluded by the invisible Higgs decay constraints.
The analytical formulas for the DM annihilation cross sections are given in appendix \ref{AnnCS}.

\section{Direct Detection}
\label{DD}
\begin{figure}
\begin{center}
\includegraphics[width=.7\textwidth,angle =0]{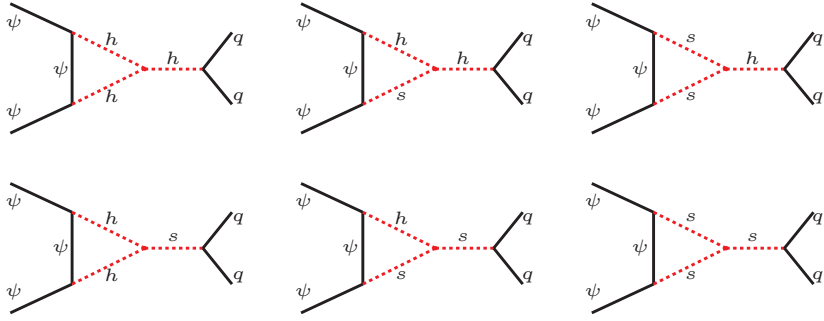}
\end{center}
\caption{The leading loop diagrams for DM Spin Independent elastic scattering off the 
 SM quarks.} 
\label{triangle}
\end{figure}

In the model we study here the DM interaction with the SM particles is of pseudoscalar type,
and at tree level its Spin Independent cross section is 
obtained in the following formula 
\begin{equation}
 \sigma^p_{\text{SI}} = \frac{2}{\pi} \frac{\mu^4 A^2}{m_{\text{dm}}^2} v_{\text{dm}}^2 \,,
\end{equation}
where $\mu$ is the reduced mass of the DM and the proton, $v_{\text{dm}} \sim 10^{-3}$ 
is the DM velocity,
and $A$ is given by 
\begin{equation}
 A = \frac{g_d \sin 2\theta}{2v_h}(\frac{1}{m_h^2}-\frac{1}{m_s^2}) \times 0.28~m_p \,,
\end{equation}
where the number $0.28$ incorporates the hadronic form factor and $m_p$ denotes the proton
mass. 
Therefore, the DM-nucleon scattering cross section is 
velocity suppressed at tree level. 
Other words, the entire parameter space of this model 
resides well below the reach of the direct detection experiments. 
The current underground DD experiments like LUX \cite{Akerib:2016vxi} 
and XENON1T \cite{Aprile:2017iyp} granted us 
with the strongest exclusion limits for DM mass 
to be in the range $\sim 10$ GeV up to $\sim 10$ TeV.
The future DD experiments can only probe direct interaction of the DM-nucleon 
down to the cross sections comparable with that of the  
neutrino background (NB), $\sigma_{\text{NB}} \sim {\cal O}(10^{-13})$ pb \cite{Billard:2013qya}. 
In the present model, as we will see in our numerical results 
the tree level DM-nucleon DD cross section is orders of 
magnitude smaller than NB cross sections. 
For such a model with the DM-nucleon cross section being velocity-suppressed 
at tree level, 
it is mandatory to go beyond tree level and find the SI cross section. 
The leading diagrams (triangle diagrams) contributing to the SI cross section are drawn
in Fig.~\ref{triangle}. There are also contributing box diagrams to the DM-nucleon
scattering process. The box diagrams bring in
a factor of $m_q^3$ ($q$ stands for light quarks) as shown in \cite{Ipek:2014gua}, 
while the triangle diagrams are proportional to $m_q$. Thus, we consider
the box diagrams to have sub-leading effects. 
We then move on to compute the leading loop effects on the SI scattering 
cross section.
In the following we write out the full expression for the DM-quark scattering 
amplitude when scalars in the triangle loop have masses $m_i$ and $m_j$ and that 
coupled to quarks has mass $m_k$,

\begin{equation}
  i{\cal M}^{ijk} = 
  \Big[\frac{C_k}{(p_1-p_2)^2-m_{k}^2} \Big] \bar q q~  \times 
\int \frac{d^4q}{(2\pi)^4}  \frac{g_d^2 ~~\bar \psi(p_{2}) \gamma^5 (\slashed{q}+m_{\text{dm}}) \gamma^5 \psi(p_1) }
{[(p_2-q)^2-m_{i}^2][(p_1-q)^2-m_{j}^2][q^2-m_{\text{dm}}^2]}   \,.
\end{equation}

In the above, the indices $i,j$ and $k$ stand for the Higgs ($h$) or the singlet scalar ($s$). 
In the expression above, we have 
$C_h= -m_q/v_h \cos \theta$ and $C_s= m_q/v_h \sin \theta$.
The corresponding effective scattering amplitude in the limit that the 
momentum transferred to a nucleon is $q^2 \sim 0$, follows this formula, 
\begin{equation}
 i{\cal M}^{ijk}_{\text{eff}} =  -i \frac{m_{\text{dm}} g_d^2}{16\pi^2} C_{ijk} F(\beta_i, \beta_j) \frac{C_k}{m_k^2} 
 ~(\bar q q)(\bar \psi \psi)   \,,
\end{equation}
in which $\beta_i = m_i^2/m_{\text{dm}}^2$ and $\beta_j = m_j^2/m_{\text{dm}}^2$, 
and the loop function $F(\beta_i, \beta_j)$ is given in appendix \ref{DDcs}. 
In the cases that the two scalar masses in the triangle loop are identical, i.e. $m_i = m_j$, 
then let's take $\beta_i = \beta_j $ and represent $F(\beta_i, \beta_j)$ by $F(\beta_i)$ 
which is provided by appendix \ref{DDcs}. The validity of these loop functions
are verified upon performing numerical integration of the Feynman integrals 
and making comparison for a few distinct input parameters. 
$C_{ijk}$ is the trilinear scalar coupling, where
there are four of them corresponding to the vertices $hhh$, $hhs$, $ssh$ and
$sss$ as appeared in Fig.~\ref{triangle}.

Putting together all the six triangle diagrams, we end up having the expression below
for the total effective SI scattering amplitude,
\begin{equation}
\begin{aligned}
 {\cal M}_{\text{eff}} {} & =   \frac{m_q}{v_h} \frac{m_{\text{dm}} g_d^2}{16\pi^2} 
 \Big[
  \frac{\cos \theta}{m_h^2} C_{hhh} F(\beta_h)  
  +\frac{\cos \theta}{m_h^2} C_{hsh} F(\beta_h, \beta_s) 
  +\frac{\cos \theta}{m_h^2} C_{ssh} F(\beta_s)   \\
  &   -\frac{\sin \theta}{m_s^2} C_{hhs} F(\beta_h)  
      -\frac{\sin \theta}{m_s^2} C_{hss} F(\beta_h, \beta_s) 
      -\frac{\sin \theta}{m_s^2} C_{sss} F(\beta_s)
      \Big] 
 ~(\bar q q)(\bar \psi \psi)  
 \\ & 
 \equiv m_q ~\alpha ~(\bar q q)(\bar \psi \psi)
 \,,
\end{aligned}
\end{equation}
The Spin Independent DM-proton scattering is 
\begin{equation}
 \sigma^p_{\text{SI}} = \frac{4\alpha_p^2\mu^2}{\pi} \,,
\end{equation}
in which $\mu$ is the reduced mass of the DM and the proton, and

\begin{equation}
 \alpha_p = m_{p} \alpha \Big( \sum_{q = u,d,s} F^{p}_{T_q}  
+ \frac{2}{9} F^{p}_{T_g}   \Big) \sim 0.28~ m_{p} \alpha \,,
\end{equation}
where $m_p$ is the proton mass and the quantities $F^{p}_{T_q}$ and $F^{p}_{T_g}$ 
define the scalar couplings for the strong interaction at low energy.
The trilinear couplings in terms of the mixing angle and 
the relevant couplings in the Lagrangian and, the DD cross section 
at tree and loop level are given in appendix \ref{DDcs}.
The scalar form factors used in our numerical computations are, 
$F^{p}_{u} = 0.0153$, $F^{p}_{d} = 0.0191$ and $F^{p}_{s} = 0.0447$ \cite{Belanger:2013oya}. 
To obtain the scalar form factors, the central values of the following 
sigma-terms are used, $\sigma_{\pi N} = 34\pm 2$ MeV and 
$\sigma_{s} = 42\pm 5$ MeV.
We computed the correction to the DD cross section at loop level 
by including the uncertainty on the two sigma-terms. 
We found that the corresponding uncertainty on the DD cross section 
are not big enough to be seen in the plots. However, we estimated 
the uncertainty for a given benchmark point with $m_{\text{dm}} \sim 732$ GeV, 
$g_{d} \sim 2.17$, $g_3 = 10$ GeV and $\sin \theta = 0.02$. 
The result is $\sigma^{p}_{loop} = (3.084 \pm 0.12)\times 10^{-10}$ pb. 

\begin{figure}
\centering
\begin{subfigure}[b]{0.55\textwidth}
\hspace{-1.8cm}
\includegraphics[width=\textwidth,angle =-90]{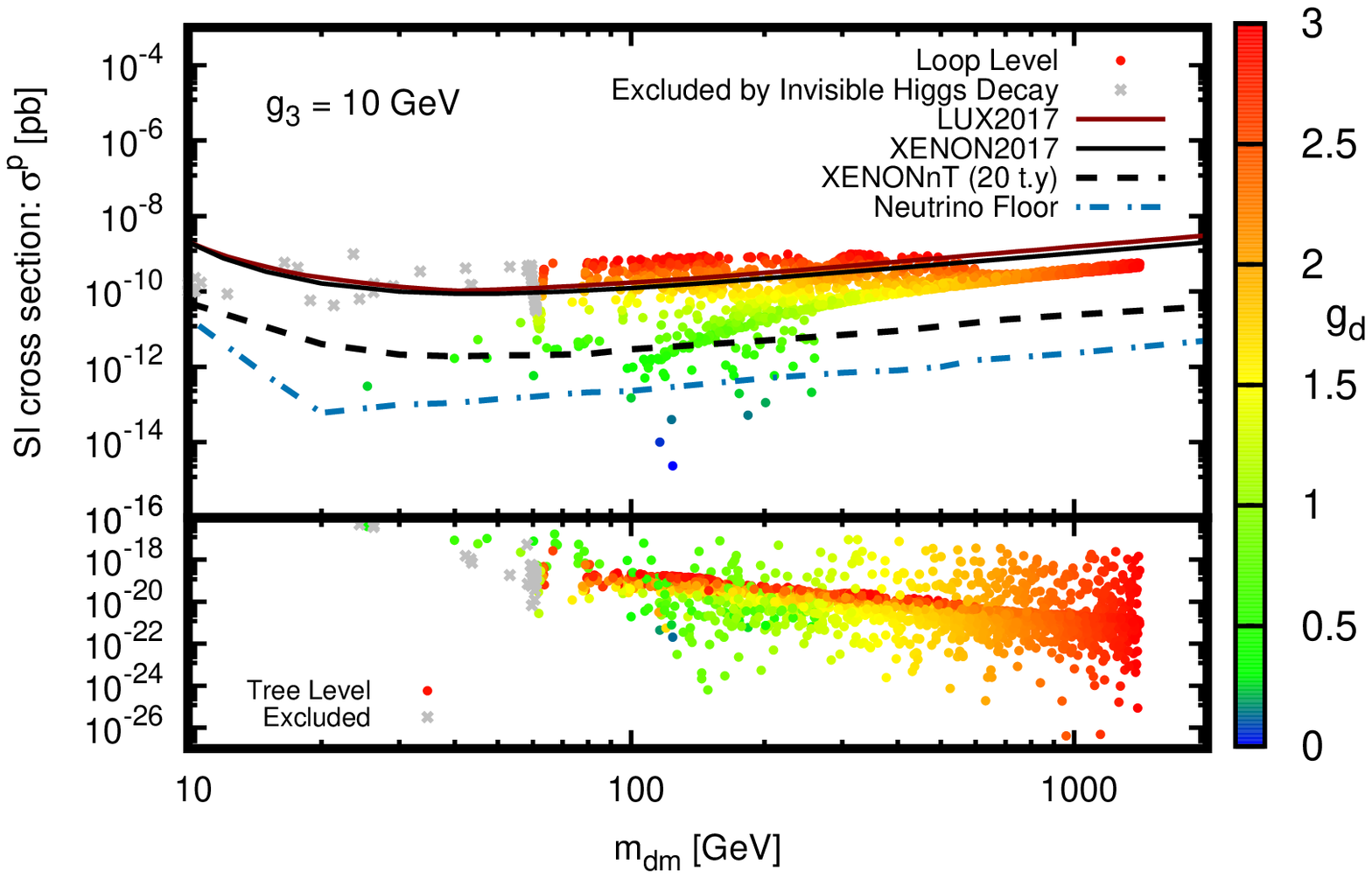}
\end{subfigure}
\begin{subfigure}[b]{0.55\textwidth}
\hspace{-1.8cm}
\includegraphics[width=\textwidth,angle =-90]{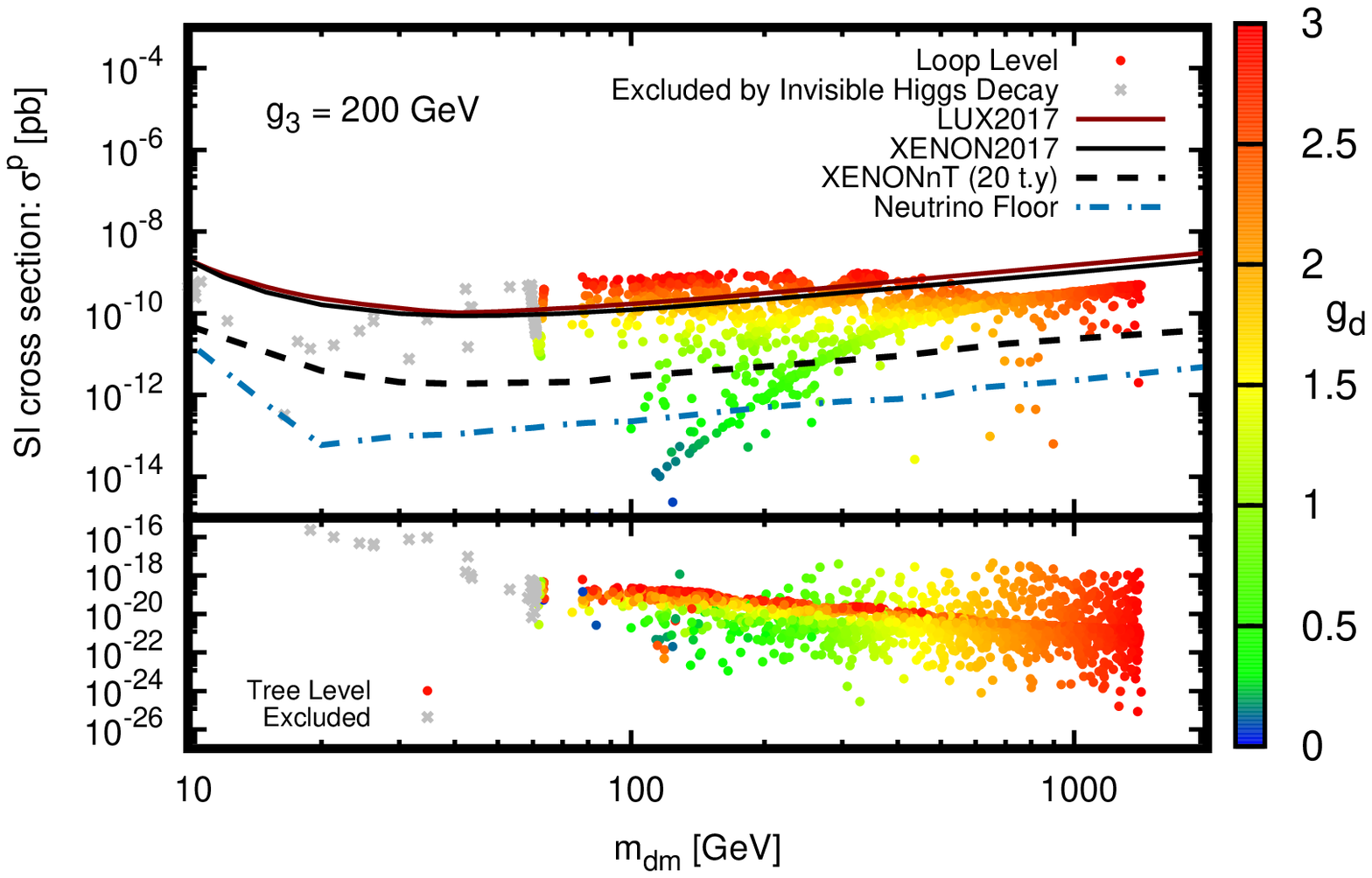}
\end{subfigure}
\caption{Shown are the DM-proton scattering cross section against the DM mass. 
In the upper panel $g_3 = 10$ GeV and in the lower panel $g_3 = 200$ GeV.
The mixing angle is such that $\sin \theta = 0.02$. The vertical color 
spectrum indicates the range of the dark coupling $g_d$. Here, the observed relic density 
constraint is applied, The upper limits from LUX and XENON1T and also XENONnT projection are shown.}
\label{direc-gs}
\end{figure}

In the first part of our scan over the parameter space we wish to compare
the DM-proton SI cross section at tree level with the SI cross section
stemming from leading loop effects. 
To this aim, we consider for the DM mass to take values as   
$10~ \text{GeV} < m_{\text{dm}} < 2~ \text{TeV}$, and the scalar mass in the range
$20~ \text{GeV} < m_s < 500~\text{GeV}$. The dark coupling varies such that
$0 < g_d < 3$.
The mixing angle in these computations is chosen a small value being $\sin \theta = 0.02$.
Reasonable values are chosen for the couplings, $g_2 = 0.1$ and $g_4 = 0.1$.
Taking into account constraints from Planck/WMAP on the DM relic density, we 
show the viable parameter space in terms of the DM mass and $g_d$
in Fig.~\ref{direc-gs} for two distinct values of the 
coupling $g_3$ fixed at $10$ GeV and $200$ GeV. Regions excluded by 
the invisible Higgs decay measurements are also shown in Fig.~\ref{direc-gs}. 
As expected the tree level SI cross section is about 10 orders of magnitude 
below the neutrino background. On the other hand, for both values of 
$g_3$, the leading loop effects are sizable in a large portion of the 
parameter space. A general feature apparent in the plots 
is that for $g_d \gtrsim 2.5$, the DM mass smaller than $600$ GeV gets 
excluded by direct detection bounds. 

\begin{figure}
\centering
\begin{subfigure}[b]{0.55\textwidth}
\hspace{-1.8cm}
\includegraphics[width=\textwidth,angle =-90]{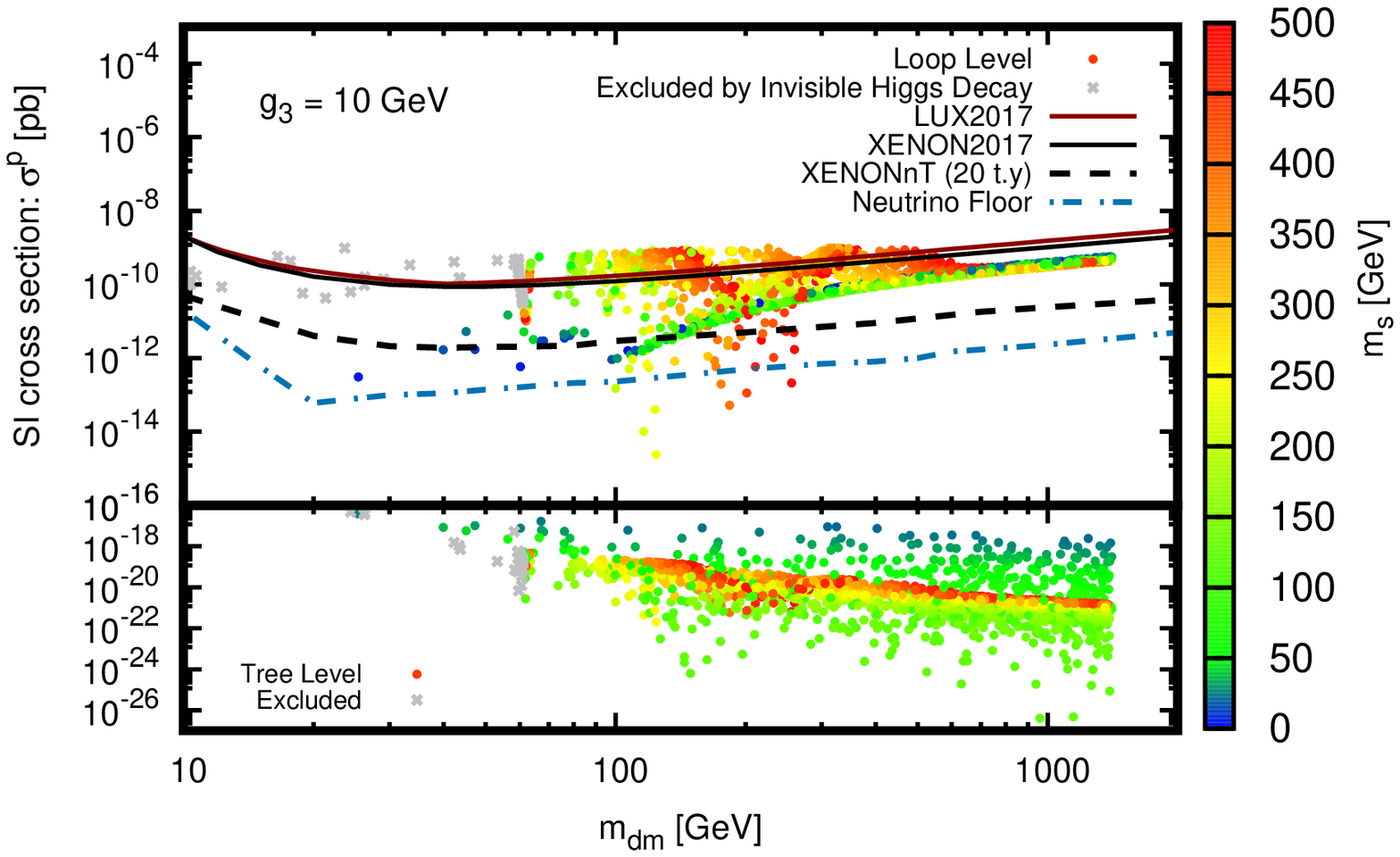}
\end{subfigure}
\begin{subfigure}[b]{0.55\textwidth}
\hspace{-1.8cm}
\includegraphics[width=\textwidth,angle =-90]{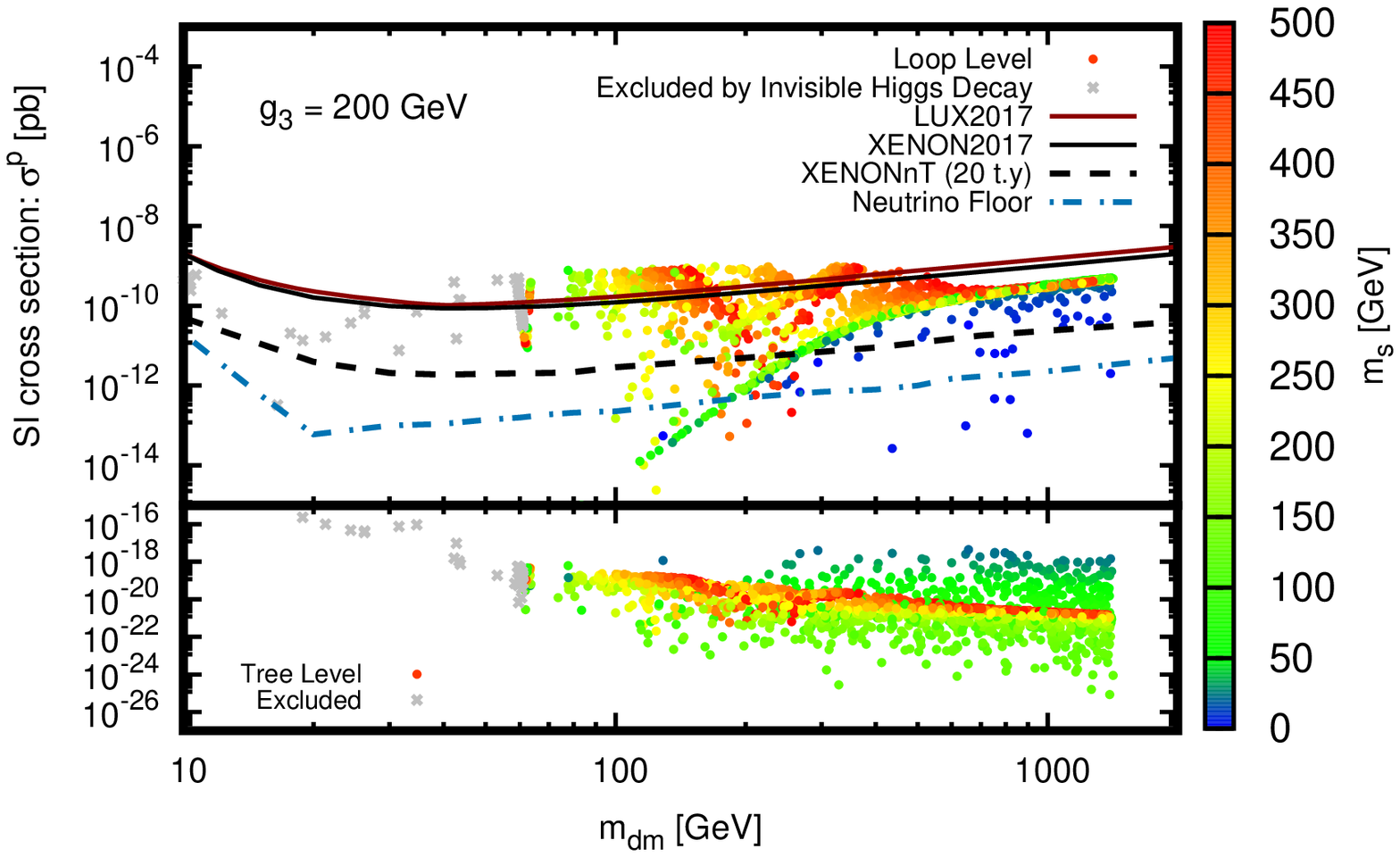}
\end{subfigure}
\caption{Shown are the DM-proton scattering cross section against the DM mass. 
In the upper panel $g_3 = 10$ GeV and in the lower panel $g_3 = 200$ GeV. 
The mixing angle is such that $\sin \theta = 0.02$.
The vertical color spectrum indicates the range of the singlet scalar mass, $m_s$. 
Here, the observed relic density constraint is applied, The upper 
limits from LUX and XENON1T and also XENONnT projection are shown.}
\label{direc-ms}
\end{figure}

In addition, with the same values in the input
parameters, we show the viable regions in terms of the DM mass and the single 
scalar mass in Fig.~\ref{direc-ms}. It is found that in both cases of
the coupling $g_3$, a wide range of the scalar mass, 
i.e, $10~\text{GeV} < m_s < 500~\text{GeV}$ 
lead to the SI cross sections above the neutrino floor. 
It is also evident from the results in Fig.~\ref{direc-ms}
that the viable region with $m_s \sim 10$ GeV located at $m_{\text{dm}} \lesssim 100$ GeV 
in the case that $g_3 = 10$ GeV, is shifted to regions with $m_{\text{dm}} \gtrsim 250$ GeV
in the case that $g_3 = 200$ GeV.

\begin{figure}
\hspace{-1.5cm}
\begin{minipage}{0.41\textwidth}
\includegraphics[width=\textwidth,angle =-90]{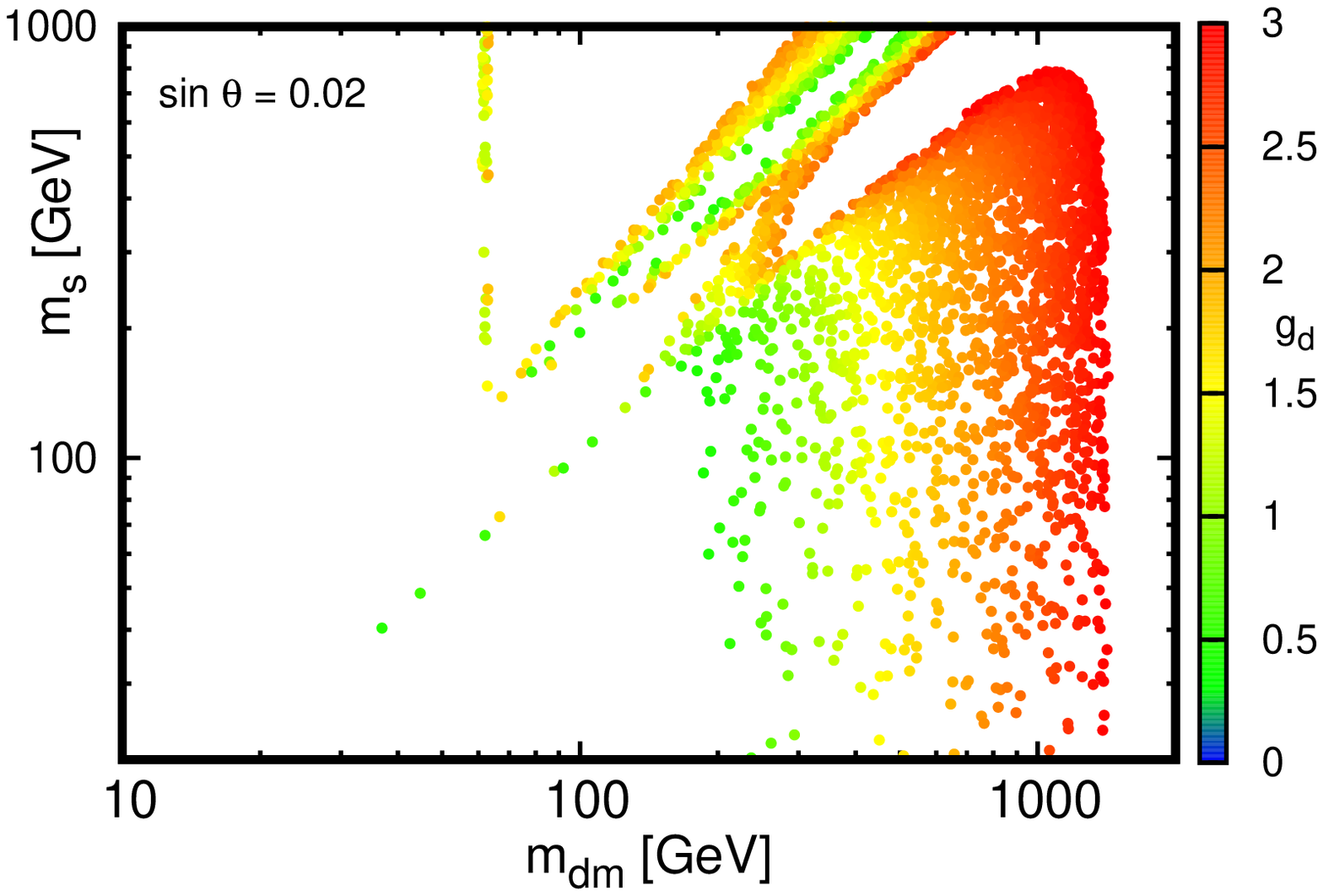}
\end{minipage}
\hspace{2.7cm}
\begin{minipage}{0.41\textwidth}
\includegraphics[width=\textwidth,angle =-90]{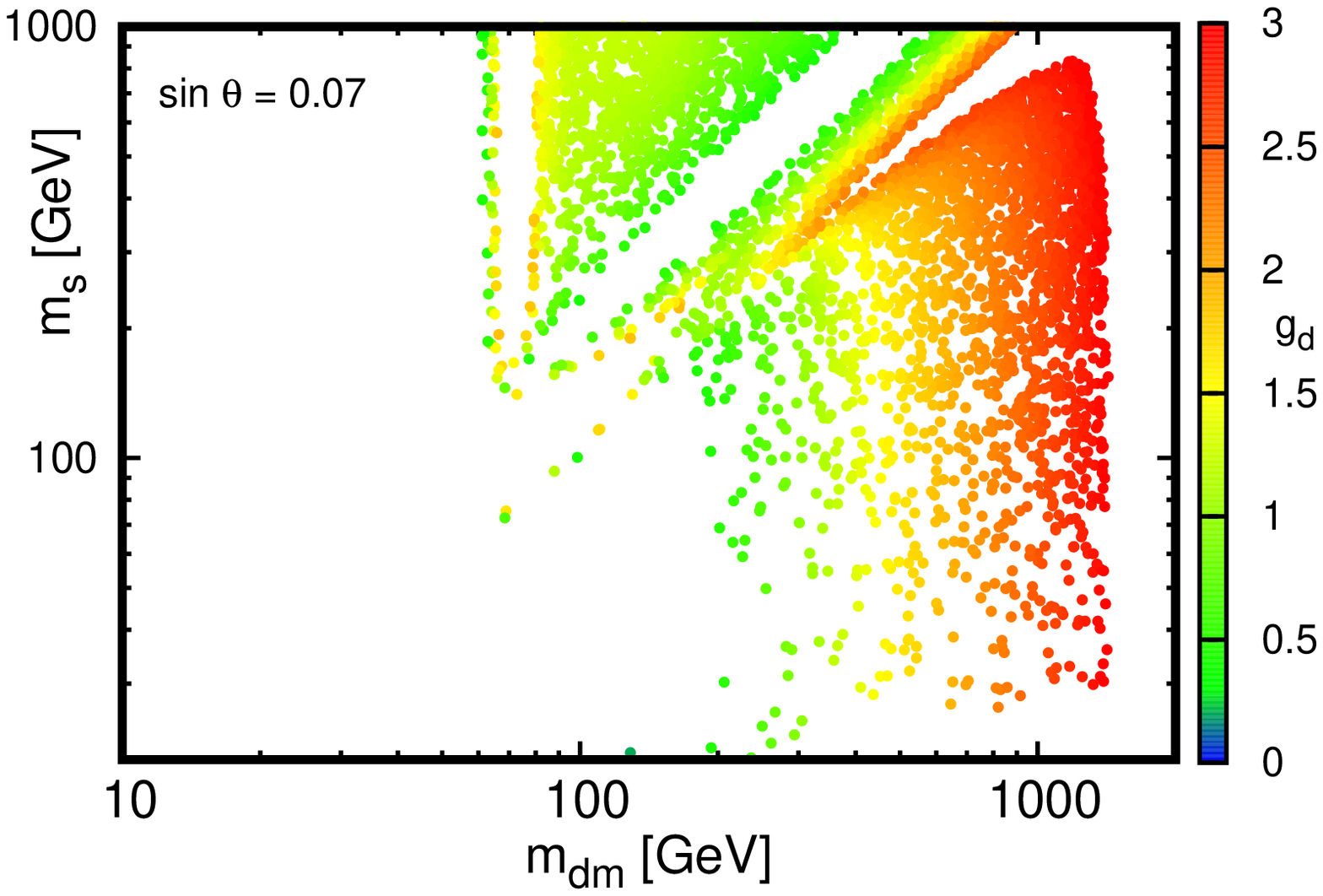}
\end{minipage}
\caption{Viable regions in the parameter space residing above the neutrino floor and below the 
current direct detection exclusion bounds are shown. Constraints from the observed relic density 
and the invisible Higgs decay are applied as well.
In the left plot, the mixing angle is such that $\sin \theta = 0.02$ and in 
the right plot $\sin \theta = 0.07$. 
In both plots, $g_3 = 200$ GeV. The vertical color spectrum indicates the range of the dark coupling, $g_d$.
}
\label{viable-region}
\end{figure}

In the last part of our computations we perform an exploratory scan 
in order to find the region of interest which are the points 
with the SI cross sections above the neutrino floor and below the DD upper limits, 
with other constraints imposed including the observed DM relic density and the invisible Higgs decay. 
The scan is done with these input parameters: 
$ 10~\text{GeV} < m_{\text{dm}} < 2~\text{TeV}$, $ 20~\text{GeV} < m_s < 1~\text{TeV}$, 
$0< g_d < 3$, $g_1=g_4 = 0.1$ and $g_3$ fixed at $200$ GeV. Our results are shown 
in Fig.~\ref{viable-region}. The mixing angle is set to $\sin\theta = 0.02$
in the left panel and $\sin\theta = 0.07$ in the right panel.    
It can be seen that for larger mixing angle the viable region is slightly 
broadened towards heavy pseudoscalar masses for the DM mass $
60~\text{GeV} < m_{\text{dm}} < 300$ GeV, also is shrank towards regions with $m_{\text{dm}} \gtrsim 60$ GeV
due to the invisible Higgs decay constraint.
We also realize that if we confine ourselves
to dark coupling $g_d \lesssim 1$ there are still 
regions with $m_{\text{dm}} \lesssim 400$ GeV which are within reach in the future 
direct detection experiments. 

Concerning indirect detection of DM, the Fermi Large Area Telescope (Fermi-LAT) 
collected gamma ray data from the Milky Way Dwarf Spheroidal Galaxies for six years \cite{Ackermann:2015zua}. 
The data indicates no significant gamma-ray excess. However, it can provide us with 
exclusion limits on the DM annihilation 
into $b\bar b$, $\tau \bar \tau$, $u\bar u$ and $W^+ W^-$ in the final state. 
As pointed out in \cite{Baek:2017vzd} the Fermi-LAT data can exclude regions in 
the parameter space with $m_{\text{dm}} < 80$ GeV and also resonant region with $m_{\text{dm}} \sim m_s/2$.

A few comments are in order on the LHC constraints beside the invisible 
Higgs decay measurements. Concerning the mono-jet search in this scenario, 
it is pointed out in \cite{Baek:2017vzd} that even in the region 
with $m_s > 2 m_{\text{dm}}$ which has the largest production rate, 
the signal rate is more than one order of 
magnitude beneath the current LHC reach, having chosen the small mixing angle. 
In the same study it is found out that bounds corresponding to di-Higgs production 
at the LHC via the process $pp \to s \to h h$, with different final 
states ($4b,2b2\gamma,2b2\tau$) are not strong enough 
to exclude the pseudoscalar mass in the relevant range
for small mixing angle as we chose in this study.

\section{Conclusions}
\label{conclusion}
We revisited a DM model whose fermionic DM candidate has a pseudoscalar interaction 
with the SM quarks at tree level leading to the suppressed SI direct detection 
elastic cross section. In the present model we obtained analytically
the leading loop diagrams contributing to the SI elastic scattering cross section.

Our numerical analysis taking into account the limits from the observed relic density, 
suggests that regions with dark coupling $g_d \gtrsim 2.5$ and reasonable values
for the other parameters, get excluded by DD upper bounds. 
It is also found that regions with $g_d \lesssim 0.25$ are excluded because they 
reside below the neutrino floor.
However, a large portion of the parameter space stands above the neutrino floor
remaining accessible in the future DD experiments such as XENONnT. 

We also found regions of the parameter space above the neutrino floor
while evading the current LUX/XENON1T DD upper limits, respecting 
the observed DM relic density and the invisible Higgs decay experimental bound. 
The viable region is slightly broadened for the moderate DM mass when $\sin\theta =0.07$
in comparison with the case when $\sin\theta = 0.02$, both at $g_3 = 200$ GeV.

\appendix

\section{Annihilation cross sections}
\label{AnnCS}
The annihilation cross sections of a DM pair into a pair of the SM fermions are as the following
\begin{equation}
\label{ff}
\begin{aligned}
\sigma_{\text{ann}} v_{\text{rel}} (\bar \psi \psi \to \bar f f) {} & = 
\frac{g_{d}^2 \sin^2 2\theta}{64 \pi} 
\Big[  \frac{1}{(s-m^{2}_{h})^2+m^{2}_{h}\Gamma^{2}_{h}}
+\frac{1}{(s-m^{2}_{s})^2+m^{2}_{s}\Gamma^{2}_{s}} 
\\ &
-\frac{2(s-m^{2}_{h})(s-m^{2}_{s})+2m_{h}m_{s}\Gamma_{h}\Gamma_{s}}
{((s-m^{2}_{h})^2+m^{2}_{h}\Gamma^{2}_{h})((s-m^{2}_{s})^2+m^{2}_{s}\Gamma^{2}_{s})} \Big] \times
\\ &
 \Big(  N_{c} \times 2 s (\frac{m_{f}}{v_{h}})^2 (1-\frac{4m^{2}_{f}}{s})^{\frac{3}{2}} \Big) \,,
\end{aligned}
\end{equation}
where the number of color charge is denoted by $N_{c}$. 
In the annihilation cross sections above the dominant contributions belong 
to the heavier final states $b \bar b$ and $t \bar t$.  
The total cross section into a pair of the gauge bosons ($W^+W^-$ and $ZZ$) in the unitary gauge is given by 
\begin{equation}
\label{ww}
\begin{aligned}
\sigma_{\text{ann}} v_{\text{rel}} (\bar \psi \psi \to W^+ W^- , ZZ) {} & = 
\frac{g_{d}^2 \sin^2 2\theta}{64 \pi} 
\Big[  \frac{1}{(s-m^{2}_{h})^2+m^{2}_{h}\Gamma^{2}_{h}} 
+\frac{1}{(s-m^{2}_{s})^2+m^{2}_{s}\Gamma^{2}_{s}} 
\\ &
-\frac{2(s-m^{2}_{h})(s-m^{2}_{s})+2m_{h}m_{s}\Gamma_{h}\Gamma_{s}}
{((s-m^{2}_{h})^2+m^{2}_{h}\Gamma^{2}_{h})((s-m^{2}_{s})^2+m^{2}_{s}\Gamma^{2}_{s})} \Big] \times
\\ &
 \Big[ (\frac{m^{2}_{W}}{v_{h}})^2(2+\frac{(s-2m^{2}_{W})^2}{4m^{4}_{W}}) (1-\frac{4m^{2}_{W}}{s})^{\frac{1}{2}}
\\ &
+\frac{1}{2}(\frac{m^{2}_{Z}}{v_{h}})^2(2+\frac{(s-2m^{2}_{Z})^2}{4m^{4}_{Z}}) (1-\frac{4m^{2}_{Z}}{s})^{\frac{1}{2}}
     \Big] \,.
\end{aligned}
\end{equation}

And finally we obtain the following expression for the DM annihilation into two higgs bosons as

\begin{equation}
\label{hh}
\begin{aligned}
\sigma_{\text{ann}} v_{\text{rel}} (\bar \psi \psi \to h h) {} & = 
\frac{g_{d}^2}{32 \pi} (1-\frac{4m^{2}_{h}}{s})^{\frac{1}{2}}
\Big[  \frac{a^2 \sin^2 \theta}{(s-m^{2}_{h})^2+m^{2}_{h}\Gamma^{2}_{h}} 
+\frac{b^2 \cos^2 \theta}{(s-m^{2}_{s})^2+m^{2}_{s}\Gamma^{2}_{s}} 
\\ &
+\frac{ a b \sin 2\theta [(s-m^{2}_{h})(s-m^{2}_{s})+m_{h}m_{s}\Gamma_{h}\Gamma_{s}]}
{((s-m^{2}_{h})^2+m^{2}_{h}\Gamma^{2}_{h})((s-m^{2}_{s})^2+m^{2}_{s}\Gamma^{2}_{s})} \Big] 
\\ &
+ \frac{g_{d}^4 \sin^4 \theta}{16 \pi s} (1-\frac{4m^{2}_{h}}{s})^{\frac{1}{2}} 
\Big[s(m^{2}_{\text{dm}}-t)+m^{2}_{\text{dm}}m^{2}_{h}
\\ &
-(m^{2}_{\text{dm}}+m^{2}_{h}-t)^2 \Big]  \times
\Big(\frac{1}{t-m^{2}_{\text{dm}}}+\frac{1}{u-m^{2}_{\text{dm}}}\Big)^2 \,,
\end{aligned}
\end{equation}
with 

\begin{equation}
\begin{aligned}
a {} & = 3 \cos^2 \theta \sin \theta g_{1} +6 \sin^2 \theta \cos \theta g_{2} v_{h}
          + 6 \cos^3 \theta \lambda v_{h} +  \sin^3 \theta g_{3} \,, \\ 
b & = 3 \cos \theta \sin^2 \theta g_{1} -  \cos \theta g_{1} + 6 \sin^3 \theta  g_{2} v_{h}
    - 4 \sin \theta g_{2} v_{h} + 6 \cos^2 \theta \sin \theta \lambda v_{h}
\,,      
\end{aligned}
\end{equation}

and DM annihilation into two $s$ bosons as

\begin{equation}
\label{ss}
\begin{aligned}
\sigma_{\text{ann}} v_{\text{rel}} (\bar \psi \psi \to s s) {} & = 
\frac{g_{d}^2}{32 \pi} (1-\frac{4m^{2}_{s}}{s})^{\frac{1}{2}}
\Big[  \frac{c^2 \sin^2 \theta}{(s-m^{2}_{h})^2+m^{2}_{h}\Gamma^{2}_{h}} 
+\frac{d^2 \cos^2 \theta}{(s-m^{2}_{s})^2+m^{2}_{s}\Gamma^{2}_{s}} 
\\ &
+\frac{c~d \sin 2\theta [ (s-m^{2}_{h})(s-m^{2}_{s})+m_{h}m_{s}\Gamma_{h}\Gamma_{s}]}
{((s-m^{2}_{h})^2+m^{2}_{h}\Gamma^{2}_{h})((s-m^{2}_{s})^2+m^{2}_{s}\Gamma^{2}_{s})} \Big] 
\\ &
+ \frac{g_{d}^4 \cos^4 \theta}{16 \pi s} (1-\frac{4m^{2}_{s}}{s})^{\frac{1}{2}} 
\Big[s(m^{2}_{\text{dm}}-t)+m^{2}_{\text{dm}}m^{2}_{s}
\\ &
-(m^{2}_{\text{dm}}+m^{2}_{s}-t)^2 \Big]  \times
\Big(\frac{1}{t-m^{2}_{\text{dm}}}+\frac{1}{u-m^{2}_{\text{dm}}}\Big)^2 \,,
\end{aligned}
\end{equation}
with 
\begin{equation}
\begin{aligned}
c {} &=   3 \sin^3 \theta g_1 - 2 \sin \theta g_1 - 6 \cos \theta \sin^2 \theta g_2 v_{h} 
          + 2 \cos \theta g_2 v_h \\
& + 6 \cos \theta \sin^2 \theta \lambda v_h + \cos^2 \theta \sin \theta g_3   \,,\\ 
d  & = 3 \cos \theta \sin^2 \theta g_1 - 6 \cos^2 \theta \sin \theta g_2 v_h - 6 \sin^3 \theta \lambda v_h
     + \cos^3 \theta g_3  \,.      
\end{aligned}
\end{equation}

The Mandelstam variables are denoted by $s$, $t$ and $u$.

\section{DD cross section at tree level and loop level}
\label{DDcs}
At {\it tree level} the DD cross section is 
\begin{equation}
\sigma^p_{\text{SI}} \sim \frac{0.28^2~m_p^6 m_{\text{dm}}^2 g_d^2 \sin^2 2\theta}{2\pi v_h^2 (m_p + m_{\text{dm}})^4} 
(\frac{1}{m_h^2}-\frac{1}{m_s^2})^2 v_{\text{dm}}^2 \,,
\end{equation}
and the DD cross section at {\it loop level} reads
\begin{equation}
\begin{aligned}
 \sigma^p_{\text{SI}} {} & \sim \frac{0.28^2~m_p^4 m_{\text{dm}}^4 g_d^4}{64\pi^5 v_h^2 (m_p + m_{\text{dm}})^2}   
  \Big|\frac{\cos \theta}{m_h^2} C_{hhh} F(\beta_h)  
  +\frac{\cos \theta}{m_h^2} C_{hsh} F(\beta_h, \beta_s) 
  +\frac{\cos \theta}{m_h^2} C_{ssh} F(\beta_s)   \\
  &   -\frac{\sin \theta}{m_s^2} C_{hhs} F(\beta_h)  
      -\frac{\sin \theta}{m_s^2} C_{hss} F(\beta_h, \beta_s) 
      -\frac{\sin \theta}{m_s^2} C_{sss} F(\beta_s) \Big|^2 \,,
\end{aligned}
\end{equation}
where, $m_p$ is the proton mass, $v_h = 246$ GeV, $\beta_h = m_h^2/m_{\text{dm}}^2$ 
and $\beta_s = m_s^2/m_{\text{dm}}^2$.
 
We present the relevant loop function in the case $\beta_i \ne \beta_j$, as
\begin{equation}
\begin{aligned}
 m_{\text{dm}}^2 F(\beta_i, \beta_j) {} & =  -\frac{1}{2} + \frac{\sqrt{\beta_j-4}}{4(\beta_j-\beta_i)}
   \beta_j^{3/2} \log \frac{\sqrt{\beta_j^2-4\beta_j}+\beta_j-2}{\sqrt{\beta_j^2-4\beta_j}-\beta_j-2}
\\
 &  -\frac{\sqrt{\beta_j-4}}{4(\beta_j-\beta_i)}
   \beta_j^{3/2} \log \frac{\sqrt{\beta_j-4}+\sqrt{\beta_j}}{\sqrt{\beta_j-4}- \sqrt{\beta_j}} 
  -(\beta_j^2-2\beta_j)\log \beta_j -(\beta_j \to \beta_i)   \,,
\end{aligned}
\end{equation}
and when $\beta_i = \beta_j = \beta$, the loop function $F$ reads,
\begin{equation}
\begin{aligned}
m_{\text{dm}}^2 F(\beta) {} & =  -\frac{1}{4} + \frac{1}{4(\beta-4)} \Big[
            2(\beta-3)\sqrt{\beta^2-4\beta} \log \frac{\sqrt{\beta-4}+\sqrt{\beta}}{\sqrt{\beta-4}-\sqrt{\beta}} 
\\ & 
 + 2(\beta^2-5\beta+4) \log \beta 
  +2(\beta-3) \log \frac{\beta^2-4\beta-(\beta-2)\sqrt{\beta^2-4\beta}}{\beta^2-4\beta+(\beta-2)\sqrt{\beta^2-4\beta}} \Big] 
  \,.      
\end{aligned}
\end{equation}
The trilinear scalar couplings are  
\begin{equation}
\begin{aligned}
C_{hhh} {}  & = -6g_2 v_h  \sin^2\theta \cos\theta- 3g_1 \cos^2\theta \sin\theta
         -6 \lambda v_h \cos^3\theta -g_3 \sin^3\theta  \,,\\
C_{hhs} {}  & = g_2 v_h  (6\sin^3\theta - 4\sin\theta)
         +3 g_1 \sin^2\theta \cos\theta - g_1 \cos\theta \\
   &+6\lambda v_h \cos^2\theta \sin\theta 
-g_3 \sin^2\theta \cos\theta \,,\\
C_{ssh} {}  & = g_2 v_h  (6\sin^2\theta \cos\theta - 2\cos\theta)
         +(2\sin\theta - 3\sin^3\theta) g_1 \\
  & -6\lambda v_h \sin^2\theta \cos\theta  
 -g_3 \cos^2\theta \sin\theta \,,\\
C_{sss} {}  & = 6g_2 v_h \cos^2\theta \sin\theta 
         -3 g_1 \sin^2\theta \cos\theta
         +6\lambda v_h \sin^3\theta -g_3 \cos^3\theta  \,.
\end{aligned}
\end{equation}

\bibliography{ref}
\bibliographystyle{utphys}

\end{document}